\documentstyle[]{fbssuppl}
\input{psfig.sty}
\psfull
\psscalefirst

\newcommand{\degrees}{\mbox{$^{\circ}$}}
\newcommand{\piz}{\mbox{$\pi^{\circ}$}}
\newcommand{\pip}{\mbox{$\pi^{+}$}}
\newcommand{\pgg}{\mbox{\rm p$(\gamma,\gamma)$}}
\newcommand{\gpi}{\mbox{$(\gamma,\pi)$}}
\newcommand{\gpz}{\mbox{\rm $(\gamma,\piz)$}}
\newcommand{\gpp}{\mbox{\rm $(\gamma,\pip)$}}
\newcommand{\pgz}{\mbox{\rm p$(\gamma,\piz)$}}
\newcommand{\pgp}{\mbox{\rm p$(\gamma,\pip)$}}
\newcommand{\ntd}{\mbox{$\rm{N} \rightarrow \Delta$}}
\newcommand{\dsp}{\mbox{$\delta_{\pi}$}}

\title{Multipole Analyses for p\gpi\ and \pgg\ 
in the region of the $P_{33}$ $\Delta$ Resonance}
\author{A.M. Sandorfi\instnr{1}, S. Hoblit\instnr{1} and J. Tonnison\instnr{2,1}}
\instlist{Physics Dept., Brookhaven National Lab, Upton, NY 11973 \and
Physics Dept., VPI \& SU, Blacksburg, VA 24061}

\begin{document}

\maketitle

\begin{abstract}
Multipole analyses of the \pgz, \pgp\ and \pgg\ reactions are carried out
using different data sets. With sufficient constraints from polarization
observables, the ratio of E2/M1 transition amplitudes for \ntd\ (EMR) appears to
be largely insensitive to differences between recent \pgz\ cross section
measurements. We deduce a current best estimate of $EMR = -(2.85 \pm 0.34 \pm 0.21)\%$.
Back angle Compton cross sections require a value for the {\em backward spin
polarizability} \dsp\ that is significantly lower than previous expectations, with a
magnitude that is coupled to the \gpi\ cross sections.
\end{abstract}

Elastic photon (Compton) scattering and pion photo-production in the energy
region of the $\rm{P}_{33}$ $\Delta(1232)$ resonance are both rich sources of Nucleon structure
information. The proton's first order scattering response is fixed by its static
properties of mass, charge, magnetic moment and spin. The leading corrections to
this {\em point} scattering come from the dynamic rearrangement of constituent charges
and spins within the proton, and are expressed in terms of six {\em polarizability}
parameters \cite{Petrun81, Ragusa93}. These fundamental properties of the proton can
be compared to QCD through, for example, the calculational techniques of Chiral
perturbation theory ($\chi$PT) \cite{BKMS94, BGM97, Holstein97}.

Although the lifetime of the $\Delta$(1232) precludes scattering measurements, the \ntd\
transition amplitudes carry structure information. While this
transition is dominantly M1 quark spin-flip, a small E2 component is expected
from interactions with pions (either in a cloud surrounding the proton
\cite{Weise87, Iachello94, SatoLee96}, or as $q\bar{q}$ exchange currents between constituent
quarks \cite{Buchmann97}). Since nucleon models differ greatly on the mechanisms used
to generate these components, the E2 and M1 transition amplitudes provide
another sensitive testing ground.

Compton scattering, pion photo-production, and pion-nucleon scattering are
related by unitarity through a common {\em S-matrix}. Below $2\pi$ threshold, $E_{\gamma} = 309$ MeV
lab, Watson's theorem requires the \gpi\ and $(\pi,\pi)$ channels to have a common
phase \cite{Watson54}, and {\em K-matrix} theory can be used to provide a consistent,
albeit model dependent, extension of this unitarity relation to higher energies
\cite{DMW91}. Once the \gpi\ multipoles are specified, the imaginary parts of the
Compton amplitudes are completely determined by unitarity and a dispersion
calculation involving integrals of the pion multipoles can be used to generate
their real parts with the only unknowns being the nucleon polarizabilities
\cite{Lvov81, L8delta97}.

At any given energy, a minimum of 8 independent observables (for each pion
charge state) are necessary to specify the 4 photo-pion helicity amplitudes
\cite{Tabakin97}. Such complete information has never been available and most
analyses have relied almost exclusively on only four, the cross section and the
three single polarization asymmetries, $\Sigma$ (linearly polarized beam), $T$ (target)
and $P$ (recoil nucleon). Very recently, we have used Compton scattering to
provide both two new constraints on the photo-pion multipoles as well as
information on the proton polarizabilities \cite{L7emr,L8delta97}. Specific
multipoles such as the very interesting isospin $\tau = 3/2$ M1 and E2 components can be
extracted from fits to a multipole expansion of the amplitude. But since such
expansions must necessarily be truncated at some point, constraints from many
observables are needed to avoid Donnachie's ambiguity of higher partial wave
strength appearing in lower partial waves, and vice versa \cite{Donnachi73}.

A new experiment at LEGS has reported cross sections and linear beam
polarization asymmetries for the \pgg, \pgz\ and \pgp\ reactions
\cite{L8compt90, L7emr}. Recent experiments at Mainz and at Bonn have also
reported results on Compton scattering and $\pi$-production \cite{COPP96, CATS96,
Beck97, Bonn94}. At energies below the $\Delta$ (for E$_{\gamma}$ less than about 270
MeV) the results from the three labs are in substantial agreement for all three
channels. However, while the Mainz Compton cross sections are in quite good
agreement with LEGS results at all overlapping energies, the LEGS \piz\ cross
sections rise above those from Mainz in the vicinity of the $\Delta$ and are about 10\%
higher at the resonance peak.  (The LEGS \pip\ cross sections also tend to be
slightly higher than those from Mainz and Bonn, but the differences in this
channel are not as pronounced.)

\begin{figure}[bt]
\psfig{width=4.75in,file=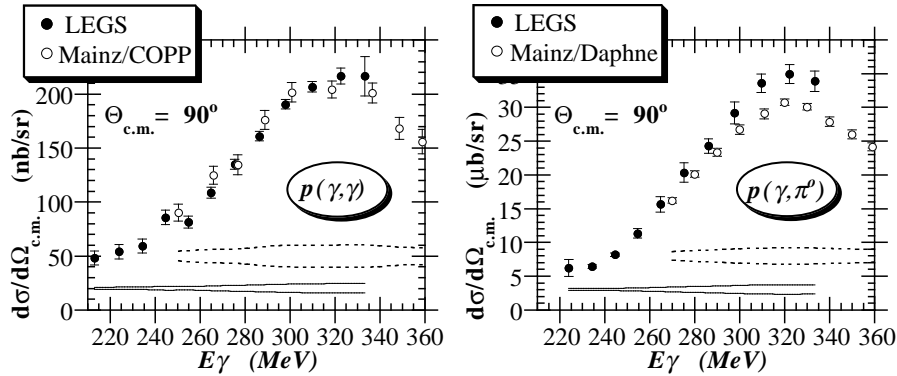}
\caption[]{Compton cross sections from LEGS \cite{L7emr} and Mainz \cite{COPP96} are
compared in the left panel, while $\pi^{\circ}$ production cross sections from \cite{L7emr}
and \cite{Beck97} are shown in the right panel. The width of the solid and dotted bands at the
bottom of the figures indicate the systematic scale uncertainties for LEGS (solid) and Mainz (dotted).}
\label{fig1}
\end{figure}

In this paper, we examine how the cross section differences among recent
$\pi$-production data sets influence the multipole decomposition of the pion
amplitude, as well as the extraction of the polarizabilities that rely on these
multipoles for the computation of dispersion integrals. 

The problem with the pion cross sections is illustrated in Fig.~\ref{fig1} where we plot the 
\pgg\ and \pgz\ results from LEGS and Mainz at 90\degrees center of mass (c.m.). The
error bars on the Mainz points are purely statistical. 
Most of the systematic effects are angle and energy dependent,
and for the LEGS results these have been evaluated point by point
and have been combined with the statistical error to produce the net uncertainty
bars. The residual systematic scale uncertainties associated with the two
measurements $(\pm\sigma_{sys})$ are indicated by bands (solid for LEGS and dotted for
Mainz). Considering these, the net accuracy of the two experiments is
comparable. Because the LEGS \pgg\ and \pgz\ measurements were made
simultaneously, in fact in the same detector, there is no possibility of independent
normalizations which could improve the agreement in the \piz\ channel without destroying the
agreement in the Compton channel.

\begin{figure}[bt]
\psfig{width=4.75in,file=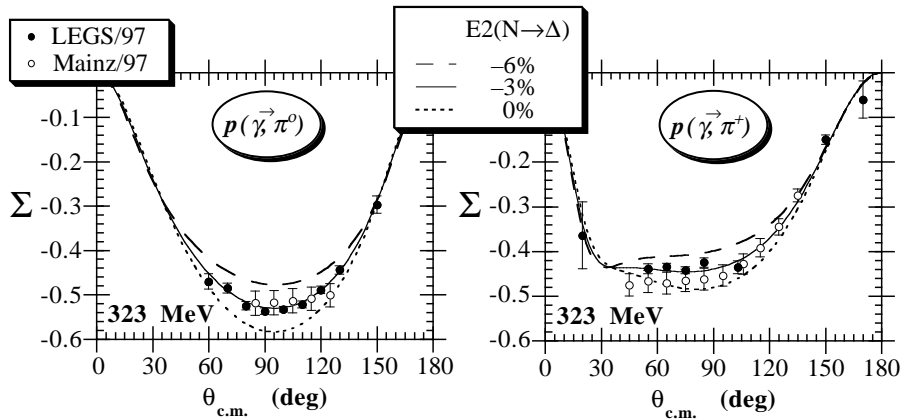}
\caption[]{Linear polarization beam asymmetries for p\gpi\ as measured at LEGS
\cite{L7emr} and Mainz \cite{Beck97, Krahn96}. Error bars reflect combined
statistical and polarization-dependent systematic uncertainties. Curves are
predictions with LEGS multipoles for different values of $\beta(E^{3/2}_{1+})$ in Eq.~(\ref{eqn1}).}
\label{fig2}
\end{figure}

The experimental agreement is better for the beam asymmetries, as shown in
Fig.~\ref{fig2}. The plotted errors include all statistical and {\em polarization-dependent}
systematic uncertainties, with polarization-independent systematic errors
canceling out of these ratios. 

To understand the extent to which these data constrain the photo-pion
multipoles, we have performed a series of energy-dependent analyses, expanding
the $\pi$-production amplitude into electric and magnetic partial waves,
$E^{\tau}_{\ell\pm}$ and $M^{\tau}_{\ell\pm}$
with relative $\pi$N angular momentum $\ell$, and intermediate-state spin $ j = \ell \pm 1/2$ and
isospin $ \tau = 1/2$ or $3/2$. The \gpi\ multipoles have been parameterized with a
K-matrix-like unitarization of the form,
\begin{equation}
A^{\tau}_{\ell\pm} = \left\{{\rm Born}_{s,u} + {\rm Born}(\rho/\omega)_{t} + {\cal P}(\alpha \cdot \varepsilon_{\pi}\right\}
\left(1 + i{\rm T}^{\ell}_{\pi N}\right) + \beta \cdot {\rm T}^{\ell}_{\pi N} .
\label{eqn1}
\end{equation}
In addition to the $s-$ and $u-$channel Born terms, and $t-$channel $\rho$ and $\omega$ exchange,
a low-order polynomial $\cal P$ in the pion energy $\varepsilon_{\pi}$ has been included to allow for
other possible terms that are expected from contributions such as $u-$channel
resonance graphs and pion rescattering \cite{L7emr}. The VPI[SM95] values have been used
for the $\pi$N scattering T-matrix elements \cite{SM95}.  Below 2$\pi$ threshold (309 MeV)
these reduce to $\sin(\delta_{\ell})e^{i\delta_{\ell}}$,
$\delta_{\ell}(E_{\gamma})$ being the elastic $\pi$N phase shift.  When a single $s-$channel
resonance dominates a partial wave having only one open decay channel the last
term in Eq.~(\ref{eqn1}) exactly reduces to a Breit-Wigner energy dependence.

Once the $(\gamma,\pi)$ multipoles are fixed by the choice of the $\alpha$ and $\beta$ parameters in
Eq.~(\ref{eqn1}) the imaginary parts of the six Compton helicity amplitudes are completely
determined by unitarity, and dispersion integrals can be used to calculate their
real parts. For the latter, we have followed the theory of L'vov \cite{Lvov81},
writing the real part of the scattering amplitude as
\begin{equation}
\Re A_{i}(\upsilon,t) = A^{B}_{i}(\upsilon,t) + \frac{2}{\pi}P\int_{\upsilon_{0}}
^{\upsilon_{max}} \frac{\upsilon' \Im A_{i}(\upsilon',t)}{\upsilon'^{2} -\upsilon^{2}}
d\upsilon' + A^{as}_{i}(t) ,
\label{eqn2}
\end{equation}
where  $\upsilon = \frac{1}{4M}(s-u)$, $M$ is the nucleon mass, and $A^{B}_{i}$ denotes the Born contribution. The
Principal value integral in (\ref{eqn2}) is calculated from $\upsilon_{0}$
(corresponding to photopion threshold) up to a moderately high energy
$(\upsilon_{max} = 1.5 GeV)$, and the $A^{as}_{i}$ are the residual asymptotic
components above $\upsilon_{max}$.

At energies below 2$\pi$ threshold, the unitarity connection between the imaginary
parts of the Compton amplitudes appearing in Eq.~(\ref{eqn2}) and the photo-pion multipoles
of Eq.~(\ref{eqn1}) is unambiguous. As $E_{\gamma}$ approaches 309 MeV, these single $\pi$-production
contributions to $\Im A_{i}$ become very large, while 2$\pi$ contributions are quite small
below 400 MeV and at higher energies are suppressed by the energy denominator in the
principle value integral of
Eq.~(\ref{eqn2}). As a result, there is in fact very little freedom in the scattering
amplitude up to the $\Delta$ peak. This allows the Compton observables to be used as an
effective constraint on the pion multipoles without incurring significant model
dependent uncertainties, provided that we restrict their use to energies below
the onset of appreciable $(\gamma,2\pi)$ strength. A reasonable set of multipoles is
needed to extend the computation of the integrals in Eq.~(\ref{eqn2}) up to 1.5 GeV, and for
this we have used VPI[SM95] \cite{SM95}, but the particular choice of the multipole
solution used for this extension has little effect on the evaluation of the
amplitudes at energies below 350 MeV. The only remaining degree of freedom in
Eq.~(\ref{eqn2}) lies in the $A^{as}_{i}$ asymptotic components. These fix the proton polarizabilities
which are determined by the $s-u = t = 0$ limits of the non-Born parts of Eq.~(\ref{eqn2}) \cite{Lvov81,L8delta97}.

With these considerations in mind, we have performed fits to the proton
polarizabilities and to the pion-multipole parameters of Eq.~(\ref{eqn1}), allowing non-Born
contributions up to $F$-waves. We have minimized $\chi^{2}$ for both the \pgg\ and the
\gpi\ observables using data in the energy region from 200 MeV to 350 MeV. When
combining data from different experiments, relative cross section normalizations
must be fitted. (To neglect this would ignore the systematic uncertainties that
are present in every experiment and thus assume an unphysical level of accuracy.) We
have followed the procedure of \cite {DAgost94}, multiplying all data from a set with
a systematic scale error $(\sigma_{sys})$ by a common factor $(f)$ while adding
$(f-1)^{2}/\sigma^{2}_{sys}$
to the $\chi^{2}$ .  The latter term weights the penalty for choosing a normalization
scale different from unity by the systematic uncertainty of the measurement.

There is good agreement among all modern Compton data and we have included in
the multipole fits all data below 350 MeV from \cite{L7emr, COPP96, CATS96,
SAL93, SAL95, MPI92, Moscow75, Ill91}. For $\pi$-production, in addition to
\pgz\ and \gpp\ cross sections and beam asymmetries, we have included in these
analyses $T$ data from \cite{BonnT96, KharTpip, KharTpi0}, $P$ data from \cite{KharTpip,
KharTpi0}, and the small amount of $G$ and $H$ beam-target double-polarization
data available from \cite{KharGH}.

Since the \gpi\ cross section differences evident in Fig.~\ref{fig1} are energy
dependent, they cannot be reconciled with a simple shift in normalization
scales. Combining them all in one multipole analysis would produce an average
result that would correspond to neither data set. Instead, we have performed
successive analyses using $\pi$-production cross sections from either LEGS, or Mainz
and Bonn, but not both simultaneously.

\begin{figure}[bt]
\psfig{width=4.75in,file=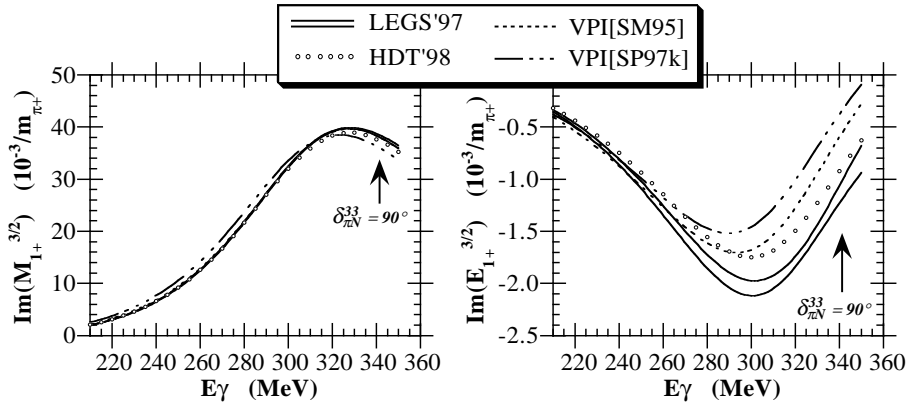}
\caption[]{The multipole solution obtained using pion $\sigma$ and $\Sigma$ data from LEGS
\cite{L7emr} is shown as solid curves which indicate the uncertainty band. For
comparison, solutions from VPI \cite{SM95} and from Mainz \cite{HDT97} are also shown.}
\label{fig3}
\end{figure}

The two most interesting results from these analysis are the M1 and E2 \ntd\
transition amplitudes and the proton polarizabilities. These quantities turn out to be
fairly decoupled, and we discuss each in turn.

\begin{table}[htb]
\caption[]{Evolution of the \ntd\ EMR, for the interval $(200 \le E_{\gamma} \le 350)$,
starting with fits to the \gpi\ cross
sections and beam asymmetries from LEGS, and expanding the data base in
subsequent rows by adding data on other observables as indicated. The number of
partial waves with fitted non-Born components is increased in successive
columns to the right.}
\label{tab1}
\begin{center}
\begin{tabular}{|ll|c|c|c|}
\hline
\multicolumn{2}{|c|}{Data included} & EMR (\%) & EMR (\%) & EMR (\%) \\
\multicolumn{2}{|c|}{successively} & $\ell_{\pi} = S - P$ &
$\ell_{\pi} = S - D$ & $\ell_{\pi} = S - F$ \\
\hline\hline
\gpi: & \{$\sigma,\Sigma$\}\cite{L7emr} & $-(2.16 \pm 0.43)$ & $-(4.22 \pm 1.08)$ & $-(4.03 \pm 1.34)$ \\
\hline
\gpi: & + \{$T$\}\cite{BonnT96} & & & \\
\multicolumn{2}{|c|}{+ \{$T,P,G,H$\}[31-33]} &
$-(2.61 \pm 0.29)$ & $-(2.74 \pm 0.28)$ & $-(2.82 \pm 0.29)$ \\
\hline
$(\gamma,\gamma)$: & + \{$\sigma,\Sigma$\}\cite{L7emr} & & & \\
& + \{$\sigma$\} {\em world}& $-(2.77 \pm 0.29)$ & $-(2.90 \pm 0.28)$ & $-(3.00 \pm 0.27)$ \\
\hline
\end{tabular}
\end{center}
\end{table}

The multipole solution obtained by taking \gpi\ cross sections and beam
asymmetries from the LEGS data \cite{L25pi0, L7emr} yields a reduced $\chi^{2}$ of
$1093/(734-36) = 1.57$. (This is the solution corresponding to row 3 of Table~\ref{tab1} in
\cite{L8delta97}.) Fitted normalization scales are all within about one standard
deviation of the systematic scale uncertainties associated with the various data
sets included. The imaginary parts of the resulting $\tau = 3/2$ $M_{1+}$ and $E_{1+}$
photo-pion multipoles are shown in Fig.~\ref{fig3} as pairs of solid lines denoting 
the uncertainty band. This is compared with two recent solutions from VPI and
one from Mainz, labeled HDT \cite{HDT97}. The LEGS, VPI[SM95] and HDT solutions are
all in agreement for the $M_{1+}$, but vary significantly in the small $E_{1+}$ 
multipole. The energy at which the $P_{33}$ phase passes through 90\degrees is indicated.
There, the LEGS and HDT solutions are fairly close. The EMR for \ntd\ is just the
ratio of the fitted  $\beta$ coefficients in Eq.~(\ref{eqn1}) for the $E^{3/2}_{1+}$ and
$M^{3/2}_{1+}$ multipoles, $-(3.00 \pm 0.27)$\%.

In a remarkably thorough but rarely quoted review article, Donnachie has pointed
out potential ambiguities that can occur when truncating a multipole expansion
\cite{Donnachi73}. These can only be mitigated by the use of many independent
observables as constraints. This is  illustrated in Table~\ref{tab1} which shows the
evolution of the fitted EMR to its final value. The number of partial waves with
fitted non-Born contributions increases to the
right in the columns while the number of observables is increased in successive
rows. If only LEGS \gpi\ cross sections and beam asymmetries \cite{L7emr} are used
as constraints (row 1), the result is unstable and strongly depends on the
number of partial waves included in the fit. But as soon as additional \gpi\  
polarization asymmetries are added (row 2), the extracted EMR value stabilizes.
Further addition of the Compton observables has only small effects (row 3). It
is by now well known that the asymmetry in \gpz\ is particularly sensitive to
the \ntd\ EMR \cite{L25pi0, Beck97}. The predictions corresponding to the -3.00\%
solution from the final analysis of Table~\ref{tab1} (the lower right-hand entry) are
shown in Fig.~\ref{fig2} as solid curves. Setting $\beta(E^{3/2}_{1+})$ to 0 or -6\% in Eq.~(\ref{eqn1})
gives the dotted and dashed curves, respectively. Despite the sizeable
separation between these curves, it should be clear from the exercises in Table~\ref{tab1}
that this observable alone is insufficient to guarantee an EMR that is free
from ambiguities.

We have also tracked the evolution of a multipole solution starting with the
Mainz \gpi\ cross sections and beam asymmetries \cite{Beck97, Krahn96}. This is
shown in Table~\ref{tab2}. When only $\sigma$ and $\Sigma$ observables are fit (row 1), the resulting
EMR again varies with the number of fitted partial waves. But when additional
polarization asymmetries are included in the fit (row 2), the EMR value
stabilizes. This is the same phenomenon encountered in Table~\ref{tab1}.

It should be noted that all of these analyses include both \gpz\ and \gpp\
data, and so are different from the treatment of \cite{Beck97} which relies on
only the \gpz\ channel. The EMR values in row 2 of Table~\ref{tab2} are smaller than
the result of \cite{Beck97}, and a contributing factor to this difference is our
inclusion of \gpp\ beam asymmetry data from \cite{Krahn96}. The centroid values of
the Mainz \gpp\ beam asymmetries tend to be more negative than the
corresponding data from LEGS. As illustrated with the calculations in Fig.~\ref{fig2},
this favors a smaller EMR. The LEGS and Mainz beam asymmetry data are in
experimental agreement (error bars from the two measurements always at least
touch), so it is appropriate to include the LEGS beam asymmetry data into this
analysis. When this is done, row 3 of Table~\ref{tab2}, the resulting EMR value doubles.
This is simply because the errors on the LEGS asymmetry data are considerably
smaller than those from the Mainz measurements and thus dominate the $\chi^{2}$ fit. The
further addition of Compton data, row 4, produces only small alterations,
although this is achieved in the fit with polarizabilities that are different
from the solution of Table~\ref{tab1}, row 3. (This is discussed further below.)

The third row of Table~\ref{tab1} and the forth row of Table~\ref{tab2} essentially agree, so
that at this point it would appear that the final \ntd\ EMR is
sufficiently constrained by the polarization asymmetry observables so as to be independent of the
\gpi\ cross section problems of Fig.~\ref{fig1}. Although this would be a highly
desirable conclusion, there is still
one complication. The Mainz data of \cite{Beck97, Krahn96} were restricted to
the $(45\degrees \leq \theta \leq 135\degrees)$ central angular range.
In this range, these data agree with
earlier measurements from Bonn \cite{Bonnpi74, Bonnpi72} that covered a much
wider angular range $(10\degrees \leq \theta \leq 180\degrees)$.
If all of these Bonn data are also included in
the fit (Table~\ref{tab2}, row 5), the resulting EMR drops by a factor of two.

The angular dependence of the differential cross section is influenced by all multipoles.
The cross sections at extreme angles are particularly sensitive to interfering multipoles of
opposite parity and can cause a significant rearrangement of multipole strength. This, and
the large number of data points from Bonn which can overwhelm a $\chi^{2}$ fit, result in the
substantial EMR shifts appearing in the last row of Table~\ref{tab2}. On the other hand, while acceptances
and efficiencies are always angle dependent, few experiments report the angular dependence of the systematic
uncertainty and none have provided the correlation of this error with angle.
Angle-independent systematic errors allow adjustments of the over all scale,
but not the shape of angular distributions, and it is the latter that affects the multipole
decomposition. For that reason, we prefer to assign the Bonn data a much lower weight in the analyses. With that
philosophy, we take the current best estimate of the \ntd\ EMR as the mean of the
$S-F$ results of Table~\ref{tab1}, row 3, and Table~\ref{tab2}, row 4. The uncertainties reported
in the tables are unbiased estimates \cite{Wolberg67} of the fitting errors that
combine {\em statistical} and {\em systematic} scale uncertainties. Additional
{\em model}-dependent errors associated with the multipole analyses have been
calculated at $\pm 0.21\%$ \cite{L7emr}. Thus, we take the current best estimate for the
EMR,
\begin{equation}
EMR = - (2.85 \pm 0.34 \pm 0.21 )\% .
\end{equation}
Unfortunately, no such mean result can be derived for the individual M1 and E2
transition amplitudes. These depend on the \gpi\ cross sections and must await a
resolution to the problems evident in Fig.~\ref{fig1}.

\begin{table}[tbh]
\caption[]{Evolution of the \ntd\ EMR, for the interval $(200 \le E_{\gamma} \le 350)$,
starting with fits to the \gpi\ cross
sections and beam asymmetries from Mainz, and expanding the data base in
subsequent rows by adding data on other observables as indicated. The number of
partial waves with fitted non-Born components is increased in successive
columns to the right.}
\label{tab2}
\begin{center}
\begin{tabular}{|ll|c|c|c|}
\hline
\multicolumn{2}{|c|}{Data included} & EMR (\%) & EMR (\%) & EMR (\%) \\
\multicolumn{2}{|c|}{successively} & $\ell_{\pi} = S - P$ &
$\ell_{\pi} = S - D$ & $\ell_{\pi} = S - F$ \\
\hline\hline
\gpi: & \{$\sigma,\Sigma$\}\cite{Beck97} & $-(1.8 \pm 0.5)$ & $-(3.4 \pm 0.8)$ & $-(2.2 \pm 0.8)$ \\
\hline
\gpi: & + \{$T$\}\cite{BonnT96} & & & \\
\multicolumn{2}{|c|}{+ \{$T,P,G,H$\}\cite{KharTpip,KharTpi0,KharGH}} &
$-(1.7 \pm 0.4)$ & $-(1.5 \pm 0.4)$ & $-(1.4 \pm 0.5)$ \\
\hline
\gpi: & + \{$\Sigma$\}\cite{L7emr} & $-(3.0 \pm 0.2)$ & $-(2.9 \pm 0.2)$ & $-(2.8 \pm 0.2)$ \\
\hline
$(\gamma,\gamma)$: & + \{$\sigma,\Sigma$\}\cite{L7emr} & $-(2.8 \pm 0.2)$ & $-(2.7 \pm 0.2)$ & $-(2.7 \pm 0.2)$ \\
\hline\hline
\gpi: & + \{$\sigma$\}\cite{Bonnpi72,Bonnpi74} & $-(1.4 \pm 0.1)$ & $-(1.5 \pm 0.1)$ & $-(1.3 \pm 0.1)$ \\
\hline
\end{tabular}
\end{center}
\end{table}

We return now to the discussion of the proton polarizabilities that are
extracted from our analyses when Compton data are included in the $\chi^{2}$
minimization, as in the solutions of Table~\ref{tab1}, row 3, and Table~\ref{tab2}, rows 4 and 5.
While four of the six Compton amplitudes of Eq.~(\ref{eqn2}) converge rapidly with
energy, the two associated with 180\degrees photon helicity-flip ($A_{1}$ and $A_{2}$) can have
appreciable asymptotic parts. In earlier analyses of data below single
$\pi$-production threshold \cite{Lvov81, SAL93, SAL95, MPI92}, $t$-channel $\pi^{\circ}$-exchange
was assumed to completely dominate $A^{as}_{2}$, which is then evaluated in terms of the
$F_{\pi^{\circ}\gamma\gamma}$ coupling. This {\em ansatz} left only $A^{as}_{1}$
to be varied in a fit to data. This
determined the difference of the electric and magnetic dipole polarizabilities,
$\bar{\alpha} - \bar{\beta}$, since the $s-u = t = 0$ limit of the $A_{1}$ amplitude is just,
\begin{equation}
\bar{\alpha} - \bar{\beta} = -\frac{1}{2\pi}A^{nB}_{1}(0,0) .
\end{equation}
Here, the $nB$ superscript denotes the non-Born contributions from the {\em integral}
and {\em asymptotic} parts of Eq.~(\ref{eqn2}). This has led to a consistent description of
Compton scattering up to $\pi$-production threshold ($E_{\gamma} \sim 150$ MeV lab), with a global
average from all data \cite{SAL95} of  $\bar{\alpha} - \bar{\beta} = 10.0 \pm 1.5{\rm (stat+sys)} \pm 0.9{\rm (model)}$, in
units of $10^{-4} fm^{3}$.

Although this had been accepted as a standard treatment of Compton scattering,
we have observed that as higher energy data were added to the fit, the deduced
value of $\bar{\alpha} - \bar{\beta}$ dropped \cite{L8delta97}, becoming even negative when Compton data up
to 2$\pi$ threshold were included from LEGS and Mainz. We have recently proposed
that the weak link here is the {\em ansatz} of no additional contributions to the
asymptotic part of the $A_{2}$ amplitude beyond those from $\pi^{\circ}$ $t$-channel exchange. We
have model corrections to $A^{as}_{2}$ with an additional exponential $t$-dependent term
having one free parameter, the derivative at $t=0$. Fitting all modern Compton
data, we have found that this addition restores consistency in $\bar{\alpha} - \bar{\beta}$ values
deduced from all data up to 2$\pi$ threshold \cite{L8delta97}.

\begin{figure}[bt]
\psfig{width=4.75in,file=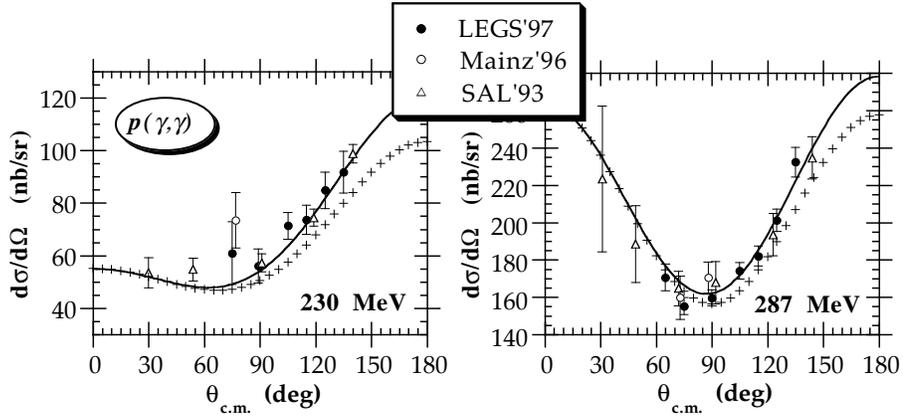}
\caption[]{Predictions with LEGS multipoles from the simultaneous fits to
\pgg\ and p\gpi\ are shown as solid curves and compared with recent data at two
energies. For these, the extracted value for $\delta_{\pi}$ is 27. Curves denoted by plus
signs used the same multipoles but held $\delta_{\pi}$ fixed at 37.}
\label{fig4}
\end{figure}

Another consequence of adding a term to $A^{as}_{2}$ is to alter the expected value for a
linear combination of the proton spin polarizabilities that characterizes
backward scattering. This {\em backward spin polarizability}, $\delta_{\pi}$, is determined by
the $s-u = t = 0$ limits of $A_{2}$ and $A_{5}$,
\begin{equation}
\delta_{\pi} = \frac{1}{2\pi M}\left[A^{nB}_{2}(0,0) + A^{nB}_{5}(0,0)\right] .
\end{equation}
Without the new variable term in $A^{as}_{2}$, the expected value for $\delta_{\pi}$ is 36.6 (in units
of $10^{-4} fm^{4}$). If $\delta_{\pi}$ is held to 36.6, the calculated Compton cross section always
falls below the back angle data. This is shown for two beam energies as curves
denoted by plus signs in Fig.~\ref{fig4}. Allowing $A^{as}_{2}$ to vary reduces $\delta_{\pi}$ and brings the
back angle predictions up in agreement with data (solid curves in Fig.~\ref{fig4}).

The fitted value of $\delta_{\pi}$ deduced from data up to $2\pi$ threshold (309 MeV) is
$27.1 \pm 2.2$ (stat+sys), with an additional model-dependent uncertainty of +2.8/-2.4
\cite{L8delta97}. (The value of $\bar{\alpha} - \bar{\beta}$ from this fit is $10.11 \pm 1.74$, in excellent
agreement with the low energy experiments.) If data up to 350 MeV are included,
as in row 3 of Table~\ref{tab1}, the deduced $\delta_{\pi}$ is $25.1 \pm 2.1$ \cite{L8delta97}. These values
for $\delta_{\pi}$ are appreciably different from the $\pi^{\circ}$-dominated expectation.

To examine the sensitivity of the deduced value of $\delta_{\pi}$ upon the \gpi\ multipole
solution we have refit the Compton data from LEGS, Mainz and SAL \cite{L7emr,COPP96,CATS96,SAL93}
using the HDT multipoles from \cite{HDT97} and the SP97k solution from
VPI \cite{SM95}. The results are listed in the first row of Table~\ref{tab3}. These two
solutions were fitted to the Mainz and Bonn \gpi\ data. If $\delta_{\pi}$ is fixed to 37, the
Compton predictions using either of these are lower than the plus-sign curves of
Fig.~\ref{fig4}. So a value for $\delta_{\pi}$ even lower than 27 is needed to raise the predictions
up to the scattering data. This is a general feature of multipoles that are fit
to the lower \gpi\ cross sections of Fig.~\ref{fig1}. Our fits in Table~\ref{tab2} give 21 for $\delta_{\pi}$ 
with the multipoles of row 4, and 19 when the Bonn \gpi\ data are included in
row 5. (The result for $\delta_{\pi}$ is almost independent of the number of partial waves,
varying by at most 2 across the columns of Table~\ref{tab2}.) 

\begin{table}[bth]
\caption[]{Results of fits to Compton data up to $2\pi$ threshold \cite{L7emr,
COPP96, CATS96, SAL93} using different \gpi\ multipoles from \cite{L8delta97,
HDT97, SM95}. In all cases, $\bar{\alpha} - \bar{\beta}$ is fixed at 10 and
$\bar{\alpha} + \bar{\beta}$ to 13.7 $(10^{-4} fm^{3})$. For
fits in the first row, the $\sigma$ mass was fixed at 600 MeV and $\delta_{\pi}$
was varied. For the second row, $\delta_{\pi}$ was fixed at 37 $(10^{-4} fm^{3})$,
and the $\sigma$ mass was varied. The $\chi^{2}$/point for all fits is less than 1.4.}
\label{tab3}
\begin{center}
\begin{tabular}{|c|c|c|c|}
\hline
& \multicolumn{3}{|c|}{\gpi\ multipoles} \\
& LEGS '98 & HDT '98 & SP97K \\
\hline
$m_{\sigma} = 600$ & $\delta = 27.1 \pm 2.2$ & $\delta = 21.4 \pm 0.9$ & $\delta = 20.9 \pm 0.8$ \\
\hline
$\delta = 37$ & $m_{\sigma} = 217 \pm 6$ & $m_{\sigma} = 82 \pm 20$ & $m_{\sigma} = 58 \pm 23$ \\
\hline
\end{tabular}
\end{center}
\end{table}

There has been a recent suggestion \cite{BGLMN98} of a possible way to fit the
Compton data while leaving the value of $\delta_{\pi}$ at its $\pi^{\circ}$-dominated expectation of
37. The asymptotic part of the A1 amplitude is assumed to be dominated by
$t$-channel $\sigma$-exchange, with $\sigma$ being the correlated s-wave $2\pi$ object required in analyses
of N-N scattering \cite{Lvov81}. Since its couplings are poorly known they are simple
treated as a free parameter in fitting $A^{as}_{1}$. In this procedure we have set the $\sigma$ 
mass to 600 MeV, an average of several N-N analyses. The authors of \cite{BGLMN98}
have pointed out that reducing $m_{\sigma}$ changes the $t$-dependence in such a way as to
raise the back angle cross section so that one might be able to reconcile
predictions with data in this way while leaving $\delta_{\pi}$ fixed at 37. We have
investigated this suggestion, and the results of refitting the Compton data,
varying $m_{\sigma}$ while fixing $\delta_{\pi} = 37$, are shown in row 2 of Table~\ref{tab3}. Good fits can
indeed be obtained in this way, but only with a value for $m_{\sigma}$ that is
substantially less than the mass of two pions. This does 
not seem a realistic alternative.

\begin{figure}[b]
\psfig{width=4.75in,file=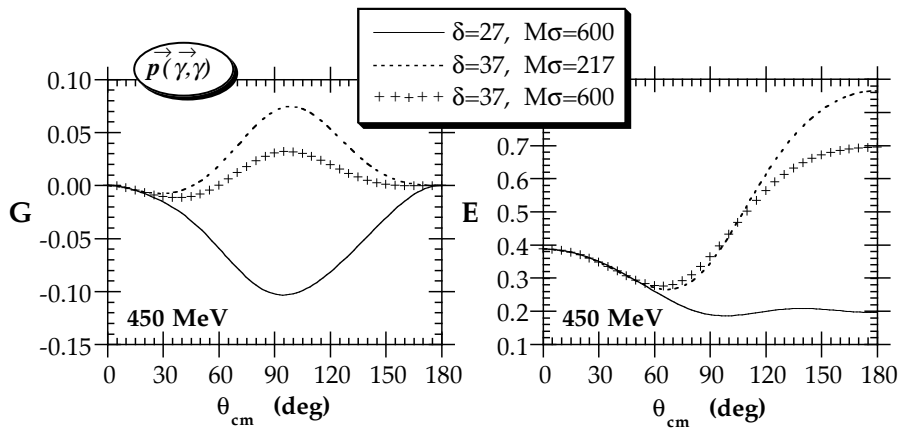}
\caption[]{Predictions from the LEGS multipoles for Compton double-polarization
observables, $G$ linearly-polarized and $E$ circularly-polarized beam, both on
longitudinally-polarized proton targets.}
\label{fig5}
\end{figure}

A value of $\delta_{\pi}$ appreciably lower than 37 is difficult to accommodate within
existing theories. Although $\chi$PT cannot be expected to directly predict Compton
observables at the high energies included in these dispersion analyses, it
should be able to reproduce the polarizabilities obtained by evaluating the
fitted amplitudes at $s-u = t = 0$. Nonetheless, existing $O(\omega^{3})$ calculations
remain close to the $\pi^{\circ}$-dominated value \cite{Holstein97}. Since our result for
$\delta_{\pi}$ would indicate some new contribution from the low-energy spin structure of the
proton, it is highly desirable to verify this in some independent way. As
pointed out in \cite{BGLMN98}, beam-target double-polarization observables are
sensitive to both $\delta_{\pi}$ and the $\sigma$ mass. In Fig.~\ref{fig5} we plot angular distributions
predicted with the LEGS multipoles for two such observables: the $G$-asymmetry
obtained with linearly polarized beam on longitudinally polarized protons (the
$\Sigma_{1Z}$ observable in \cite{BGLMN98}), and the $E$-asymmetry from circularly polarized
beam on a longitudinally polarized target ($\Sigma_{2Z}$  in \cite{BGLMN98}). Checking either
the $\delta_{\pi}$ = 37, or reduced $m_{\sigma}$ predictions should be quite straight forward, and
measurements of these quantities are expected in the near future.
(Unfortunately, the large sensitivities evident in Fig.~\ref{fig5} only occur for
energies above the $P_{33}$ resonance. Since this is now closer to the onset of
appreciable $(\gamma,2\pi)$ strength, the model dependence will increase. But one should
be able to estimate this effect using data on both $E$ and $G$.) Since the value of $\delta_{\pi}$ 
does depend upon the \gpi\ cross sections, constraining $\delta_{\pi}$ will in turn increase
the effectiveness of Compton scattering as a constraint on $\pi$-production.

\begin{acknowledge}
We are indebted to Dr. B. Preedom for his critical comments.
This work was supported by the U.S. Department of Energy under contract No.
DE-AC02-98CH10886, and by the National Science Foundation.
\end{acknowledge}

\makeatletter \if@amssymbols%
\clearpage 
\else\relax\fi\makeatother

\SaveFinalPage

\begin{thebibliography}{99}
\bibitem{Petrun81} V.A. PetrunÕkin, Sov. J. Part. Nucl. {\bf 12}, 278 (1981).
\bibitem{Ragusa93} S. Ragusa, Phys. Rev. {\bf D47}, 3757 (1993).
\bibitem{BKMS94} V. Bernard, N. Kaiser, Ulf-G. Mei{\ss}ner and A. Schmidt, Z. Phys.
A348, 317 (1994).
\bibitem{BGM97} D. Babusci, G. Giordano and G. Matone, Phys. Rev. {\bf C55}, R1645 (1997). 
\bibitem{Holstein97} T. Hemmert, B. Holstein, J. Kambor and G. Kn\"{o}chlein,
Phys. Rev. {\bf D57}, 5746 (1998).
\bibitem{Weise87} A. Wirzba and W. Weise, Phys. Lett. {\bf B188}, 6 (1987).
\bibitem{Iachello94} R. Bijker, F. Iachello and A. Leviatan, Ann. Phys. {\bf 236}, 69 (1994).
\bibitem{SatoLee96} T. Sato and T.-S.H. Lee, Phys. Rev. {\bf C54}, 2660 (1996).
\bibitem{Buchmann97} A.J. Buchmann, E. Hernndez and A. Faessler, Phys. Rev. {\bf C55}, 1 (1997).
\bibitem{Watson54} K. Watson, Phys. Rev. {\bf 95}, 228 (1954).
\bibitem{DMW91} R. Davidson, N. Mukhopadhyay, R. Wittman, Phys. Rev. {\bf D43}, 71 (1991).
\bibitem{Lvov81} A.I. L'vov, Sov. J. Nucl. Phys. {\bf 34}, 597 (1981); A.I. LÕvov, {\it et al.},
Phys. Rev. {\bf C55}, 359 (1997).
\bibitem{L8delta97} J. Tonnison, A.M. Sandorfi, S. Hoblit and A.M. Nathan, Phys.
Rev. Lett. {\bf 80}, 4382 (1998).
\bibitem{Tabakin97} W. Chiang and F. Tabakin, Phys. Rev. {\bf C55}, 2054 (1997).
\bibitem{Donnachi73} A. Donnachie, Rep. Prog. Phys. {\bf 36}, 695 (1973).
\bibitem{L8compt90} LEGS Collaboration, G. Blanpied et. al., Phys. Rev. Lett. {\bf 76}, 1023-26 (1996).
\bibitem{L7emr}	LEGS Collaboration, G. Blanpied et. al., Phys. Rev. Lett. {\bf 79}, 4337 (1997).
\bibitem{COPP96} C. Molinari et al., Phys. Lett. {\bf B371}, 181 (1996).
\bibitem{CATS96} J. Peise et al., Phys. Lett. {\bf B384}, 37 (1996).
\bibitem{Beck97} R. Beck et al., Phys. Rev. Lett. {\bf 78}, 606, (1997).
\bibitem{Bonn94} K. B\"{u}chler, et al., Nucl. Phys. {\bf A570}, 580 (1994).
\bibitem{Krahn96} H.-P. Krahn, thesis, U. Mainz (1996); R. Beck, private comm.
\bibitem{SM95} {\em SAID} code; R. Arndt, I. Strakovsky and R. Workman, Phys. Rev.
{\bf C53}, 430 (1996).
\bibitem{DAgost94} G. D'Agostini, Nucl. Inst. Meth. Phys. Res. {\bf A346}, 306 (1994).
\bibitem{SAL93} E.L. Hallin et al., Phys. Rev. {\bf C48}, 1497 (1993).
\bibitem{SAL95}	B.E. MacGibbon et al., Phys. Rev. {\bf C52}, 2097 (1995).
\bibitem{MPI92} A. Zieger et al., Phys. Lett. {\bf B278}, 34 (1992).
\bibitem{Moscow75} P.S. Baranov et al., Sov. J. Nucl. Phys. {\bf 21}, 355 (1975); only
90\degrees data are included \cite{L8delta97}.
\bibitem{Ill91} F.J. Federspiel et al., Phys. Rev. Lett. {\bf 67}, 1511 (1991).
\bibitem{BonnT96} H. Dutz et al., Nucl. Phys. {\bf A601}, 319 (1996); Gisela Anton, priv. comm.
\bibitem{KharTpip} V.A. Get'man et al., Nucl. Phys. {\bf B188}, 397 (1981).
\bibitem{KharTpi0} A. Belyaev et al., Nucl. Phys. {\bf B213}, 201 (1983).
\bibitem{KharGH} A.A. Belyaev et al., Sov. J. Nucl. Phys. {\bf 40}, 83 (1984);  {\it ibid},
43, 947 (1986).
\bibitem{L25pi0} LEGS Collaboration, G. Blanpied {\it et al.}, Phys. Rev. Lett. {\bf 69}, 1880 (1992).
\bibitem{HDT97}	O. Hanstein, D. Drechsel and L. Tiator, Nucl. Phys. {\bf A632}, 561 (1998).
\bibitem{Bonnpi74} H. Genzel et al., Z. Physik {\bf A268}, 43 (1974).
\bibitem{Bonnpi72} G. Fischer et al., Z. Physik {\bf 253}, 38 (1972).
\bibitem{Wolberg67} J.R. Wolberg, {\em "Prediction Analysis"}, Van Nostrand Co., NY, p. 54-66 (1967).
\bibitem{BGLMN98} D. Babusci, G. Giordano, A. LÕvov, G. Matone and A. Nathan, Phys. Rev. {\bf C58}, 1013 (1998).
\end{thebibliography}
\end{document}